\documentstyle[12pt]{article}
\begin{document}
\title{\bf A Normal Coordinate Expansion \\
 for the Gauge Potential \thanks{gr-qc/9507003}}
\author{F.A. Dilkes \thanks{email: fad@apmaths.uwo.ca}\\
Department of Applied Mathematics\\
University of Western Ontario\\
London ~CANADA\\
N6A 5B7}
\date{03 July 1995 \\ Revised 27 July 1995}
\maketitle
\vfill
\begin{abstract}
In this pedagogical note, I present a method for constructing
a fully covariant normal coordinate expansion of the gauge
potential on a curved space-time manifold.  Although the
content of this paper is elementary, the results may prove
useful in some applications and have not, to the best of
my knowledge, been discussed explicitly in the literature.
\end{abstract}
\section{Introduction}

Riemannian normal coordinates are first introduced in the
geometric interpretation of gravitation as a realization
of the equivalence principal which requires the existence
of an inertial reference frame at every point in
space-time in which the effects of gravity can be locally
neglected.

In addition to their axiomatic significance, normal coordinates
have found a very useful place in perturbative quantum field
theory on curved manifolds, and perhaps, in quantum gravity.
Generally covariant Taylor-type expansions in
normal coordinates have proved useful in the
path-integral environment, for example, to perform loop
calculations for the non-linear sigma model \cite{Alvarez,Honerkamp}
and to analyze trace anomalies \cite{Bastianelli}.
Generally, one can use these normal coordinate expansions to
perform the so-called proper time heat kernel expansion
(sometimes known as the DeWitt expansion \cite{DeWitt}) both
with \cite{QMPIDeWitt} and without \cite{Luscher,OConnor} the
benefit of worldline path-integral methods.  Within the framework
of the background field method, the latter expansions can be very
useful for straightforward determinations of the ultraviolet
behaviour of radiative corrections \cite{ORDeWitt}.

A good discussion of the generally covariant normal coordinate
expansion for a tensor field can be found in
reference~\cite{Alvarez} where the expansion is given
explicitly to fourth order.  However, as far as
the author is aware, very little has been published
concerning the fully covariant expansion of a gauge field.
The aim of this paper is to provide such a covariant expansion.

The following section reviews the construction of
normal coordinates, and introduces the notation which
will be used thereafter.  In section~\ref{sectGauge}
the gauge field is introduced, along with the so-called {\it radial
gauge} condition which turns out to fit very well in the
normal coordinate system and can be used to construct a generally
gauge-covariant normal coordinate expansion.

\section{Normal Coordinates}
\label{sectNormal}

Suppose we have some curved space-time manifold with a local
coordinate system $q^\alpha$ defined in the neighbourhood of
a fixed point $\phi$, and a corresponding metric
with affine connection
$\Gamma^\alpha_{\beta\gamma}(q)$.  We would like to define
what is meant by a normal coordinate system with $\phi$ at
the origin.

For any given point $q^\alpha$ we construct a geodesic
$\lambda^\alpha(q,t)$ which connects $q$ with $\phi$.  Then
$\lambda$ can be taken to satisfy the equation
\begin{equation}
\label{geodesic}
\ddot{\lambda}^\alpha(q,t) + \Gamma^{\alpha}_{\beta\gamma}
\dot{\lambda}^\beta(q,t) \dot{\lambda}^\gamma(q,t) = 0 \, ,
\end{equation}
for $t \in [0,1]$, with end-points
\begin{eqnarray*}
\lambda^\alpha(q,0) & = & \phi^\alpha \\
\lambda^\alpha(q,1) & = & q^\alpha  \, .
\end{eqnarray*}

The {\it normal coordinates} of any point $q$ are defined to be the
components of the tangent vector $\xi(q)$, at the origin $\phi$,
of the geodesic ending at $q$, i.e.
\begin{equation}
\label{normal}
\xi^\alpha(q) = \dot{\lambda}^\alpha(q,0) \, .
\end{equation}
Despite the suggestive notation, $\xi(q)$ is not a vector field
since the right hand side of equation~(\ref{normal}) is a tangent
vector at the origin $\phi$, not at $q$.

\section{The Gauge Field $ \;\;\;\; A_\alpha = A_\alpha^a T^a$ }
\label{sectGauge}

We turn our attention to the task at hand, that of developing
a fully covariant normal coordinate expansion for the gauge
potential.

To start, it has been shown~\cite{OConnor,Schmidt} that an
appropriate gauge condition for this type of expansion is the
{\it radial} or {\it synchronous gauge} (a curved-space
generalization of the Fock-Schwinger gauge~\cite{Cronstrom,Fock})
which fits very well
in the normal coordinate construction.  In the basis of the normal
coordinate system, the gauge condition is
\begin{equation}
\label{gauge}
\xi^\alpha A_\alpha(q(\xi)) = 0 \, .
\end{equation}
This condition fixes the gauge relative to a global gauge
transformation.

Using either a method of integration along the
geodesics (which is formally identical to the flat-space
problem~\cite{Schmidt,Cronstrom}) or differential forms~\cite{OConnor}
one can show, in the normal coordinate system, that the radial
gauge leads to gauge-covariant expression for the gauge potential,
\begin{equation}
\label{AF}
A_\gamma(q(\xi)) = \sum^\infty_{n=0} \frac{\left[ \xi^\beta
\nabla_\beta
(\phi) \right]^n}{n! (n+2)} \xi^\alpha F_{\alpha\gamma}(\phi) \, ,
\end{equation}
where $\nabla_\beta = \partial_\beta + {\rm adj}{A_\beta}$
is the gauge-covariant derivative in normal coordinates, and
the covariant field strength tensor is
\mbox{$F_{\alpha\gamma} = [ \nabla_\alpha , \nabla_\gamma ]$}.

Now, this latter equation is not exactly of the desired form since
the derivatives ($\nabla_\beta$) are not covariant under
reparametrization
of the manifold.  However, using the methods of
reference~\cite{Alvarez} it is straightforward to write these
gauge-covariant derivatives at the origin in terms of the
corresponding fully-covariant derivatives, denoted by indices
trailing the semicolon ($;$).  For example, one can show that
\footnote{The author suspects that the fourth derivative of
a rank-two tensor implied in reference~\cite{Alvarez} is not
entirely correct.  The corresponding coefficients presented here
for the field strength (equation~(\ref{DDDDF}))
should be reliable.\label{correction}}
\begin{eqnarray}
\nabla_{\beta_1} F_{{\beta_0}\gamma} & \dot = &
F_{{\beta_0}\gamma;\beta_1}
\label{DF} \\
\nabla_{\beta_2} \nabla_{\beta_1} F_{{\beta_0}\gamma} & \dot = &
F_{{\beta_0}\gamma;\beta_1\beta_2}
+ \frac{1}{3} R^\delta_{\beta_1 \beta_2 \gamma} F_{{\beta_0}\delta}
\label{DDF} \\
\nabla_{\beta_3} \nabla_{\beta_2} \nabla_{\beta_1}
F_{{\beta_0}\gamma}
& \dot = &
F_{{\beta_0}\gamma;\beta_1\beta_2\beta_3}
+ \frac{1}{2} R^\delta_{\beta_1\beta_2\gamma;\beta_3}
F_{{\beta_0}\delta} + R^\delta_{\beta_1 \beta_2 \gamma}
F_{{\beta_0}\delta;\beta_3}
\label{DDDF} \\
\nabla_{\beta_4} \nabla_{\beta_3} \nabla_{\beta_2} \nabla_{\beta_1}
F_{{\beta_0}\gamma} & \dot = &
F_{{\beta_0}\gamma;\beta_1\beta_2\beta_3\beta_4}
+ \frac{3}{5} R^\delta_{\beta_1\beta_2\gamma;\beta_3\beta_4}
F_{{\beta_0}\delta}
+ 2 R^\delta_{\beta_1\beta_2\gamma;\beta_3}
F_{{\beta_0}\delta;\beta_4} \nonumber \\
& & \;\;\;\;\;\;\;
+ 2 R^\delta_{\beta_1\beta_2\gamma}
F_{{\beta_0}\delta;\beta_3\beta_4}
+ \frac{1}{5} R^\delta_{\beta_1\beta_2\gamma}
R^\epsilon_{\beta_3\beta_4\delta} F_{{\beta_0}\epsilon}
{}.
\label{DDDDF}
\end{eqnarray}
where $\dot =$ indicates equality at the origin
only after symmetrization of the $\beta_i$ indices.  Substitution
of equations~\mbox{(\ref{DF})-(\ref{DDDDF})} into eq.~(\ref{AF})
yields the fully covariant normal coordinate expansion to fifth-order
in the normal coordinates,
\begin{eqnarray}
\label{AexpansionF}
A_\alpha(q) & = &
\frac{1}{2}\left\{ F_{\beta\alpha} \right\} \xi^\beta
+ \frac{1}{3} \left\{ F_{{\beta_0}\gamma;\beta_1}
\right\} \xi^{\beta_0} \xi^{\beta_1}
+ \frac{1}{8} \left\{F_{{\beta_0}\gamma;\beta_1\beta_2}
+ \frac{1}{3} R^\delta_{\beta_1 \beta_2 \gamma} F_{{\beta_0}\delta}
\right\} \xi^{\beta_0} \xi^{\beta_1}
\xi^{\beta_2} \nonumber \\
& & + \frac{1}{3! 5}
\left\{F_{{\beta_0}\gamma;\beta_1\beta_2\beta_3}
+ \frac{1}{2} R^\delta_{\beta_1\beta_2\gamma;\beta_3}
F_{{\beta_0}\delta} + R^\delta_{\beta_1 \beta_2 \gamma}
F_{{\beta_0}\delta;\beta_3}
\right\}
\xi^{\beta_0} \xi^{\beta_1} \xi^{\beta_2} \xi^{\beta_3} \nonumber \\
& & + \frac{1}{4! 6}
\left\{F_{{\beta_0}\gamma;\beta_1\beta_2\beta_3\beta_4}
+ \frac{3}{5} R^\delta_{\beta_1\beta_2\gamma;\beta_3\beta_4}
F_{{\beta_0}\delta}
+ 2 R^\delta_{\beta_1\beta_2\gamma;\beta_3}
F_{{\beta_0}\delta;\beta_4} \right. \\
& & \left. \;\;\;\;\;\;\;\;\;\;
+ 2 R^\delta_{\beta_1\beta_2\gamma}
F_{{\beta_0}\delta;\beta_3\beta_4}
+ \frac{1}{5} R^\delta_{\beta_1\beta_2\gamma}
R^\epsilon_{\beta_3\beta_4\delta} F_{{\beta_0}\epsilon}
\right\}
\xi^{\beta_0} \xi^{\beta_1} \xi^{\beta_2} \xi^{\beta_3} \xi^{\beta_4}
\nonumber \\
& & + {\cal O} \xi^6 \, .\nonumber
\end{eqnarray}
All coefficients in braces $\{ \cdots \}$ are evaluated at the
origin, where the basis vectors for the normal coordinate system
coincide with those of the original system.
Since potential on the left hand side of this equation is not
a vector at the origin, its indices must refer to the
normal coordinate basis.
(This is also true of the expansions presented in
reference~\cite{Alvarez}).
Higher order corrections to this expansion should be very
straightforward to obtain by using the methods of
ref.~\cite{Alvarez} along with equation~(\ref{AF}).

\section{Discussion}

By fusing the ungauged developments of ref.~\cite{Alvarez} with the
gauge expansion of refs.~\cite{OConnor,Schmidt},
we have been able to show how a normal coordinate expansion can
be constructed for the gauge field with fully covariant
coefficients in curved space.

Although this type of expansion may have additional uses, the
motivation for this work was in constructing a proper time (DeWitt)
expansion for the gauged heat kernel in curved space-time.  In
that context, these results could contribute to a covariant
analysis of the renormalization group in non-Abelian gauge theory.
The results of the former study are forthcoming~\cite{QMPIDeWitt}.

Finally, it should be pointed out that equation (\ref{AexpansionF})
agrees with a third-order normal
coordinate expansion which can be found in ref.~\cite{Luscher} (also
in
the context of proper time expansions).  Regrettably, the gauge
field expansion falls outside of the main interests of that work, so
all relevant details have been omitted.  The author is not aware of
anywhere else where this expansion has been constructed explicitly.

\section{Acknowledgments}

I would like to thank D.G.C.~McKeon for invaluable discussions, and
for an independent check concerning footnote~\ref{correction} on
page~\pageref{correction}.  I am also grateful for D.J.~O'Connor for
providing reference~\cite{OConnor}.

The Natural Science and Engineering Research Council of Canada
(NSERC) is acknowledged for financial support.
\eject


\begin{thebibliography}{99}
\bibitem{Alvarez} L. Alvarez-Gaum\'{e}, D.Z. Freedman, S. Mukhi,
Ann. Phys. {\bf 134}, 85-109 (1981).
\bibitem{Honerkamp} J. Honerkamp, Nucl. Phys. {\bf B36} (1972)
130-140.
\bibitem{Bastianelli} F. Bastianelli, Nucl. Phys. {\bf B376} (1992)
113-126, \mbox{hep-th/9112035}. \\
F. Bastianelli and P. van Nieuwenhuizen, Nucl. Phys. {\bf B389}
(1993) 53-80, \mbox{hep-th/9208059}.
\bibitem{DeWitt} B. DeWitt, {\it Dynamical Theory of Groups and
Fields}, Gordon and Breach (New York 1965). \\
R.T. Seeley, Amer. Math. Soc. {\bf 10}, 228 (1967). \\
P.B. Gilkey, J. Diff. Geom. {\bf 10}, 601 (1975).
\bibitem{QMPIDeWitt} F.A. Dilkes and D.G.C. McKeon, (work in
progress).
\bibitem{Luscher} M. L\"uscher, Ann. Phys. {\bf 142}, 359-392 (1982).
\bibitem{OConnor} D.J. O'Connor, PhD dissertation (University of
Maryland), 1985. \\
D.J. O'Connor, Nucl. Phys. {\bf B298}, (1988) 429-444.
\bibitem{ORDeWitt}
L. Culumovic, D.G.C. McKeon and T.N. Sherry, Ann. Phys. {\bf 197},
94 (1989). \\
L. Culumovic and D.G.C. McKeon, Can. J. Phys. {\bf 68}, 1166 (1990).
\\
D.G.C. McKeon and S.K. Wong, Int. J. Mod. Phys. A
(in press). \\
F.A. Dilkes and D.G.C. McKeon, Phys. Rev. D (to be published),
hep-th/9502075.
\bibitem{Schmidt} M.G. Schmidt and C. Schubert, Phys. Lett. {\bf
B318} (1993) 438, \mbox{hep-th/9309055}.
\bibitem{Cronstrom} C. Cronstr\"om, Phys. Lett. {\bf 90B} (1980) 267.
\bibitem{Fock} A. Fock, Phys. Z. Sowjetunion {\bf 12}, 404 (1937).
\end{thebibliography}
\end{document}